\documentclass[11pt]{article}

\usepackage{array}
\newcolumntype{P}[1]{>{\centering\arraybackslash}p{#1}}

\usepackage{booktabs} % move this to preamble and uncomment

\usepackage{hyperref}

\usepackage{tikz}
\usepackage{verbatim}

\usepackage{amssymb}
\usepackage{lineno}
\usepackage{mathtools,bbm}
\usepackage{graphicx,url}
\usepackage{graphics}

\usepackage{pdfpages}

\DeclareMathOperator*{\argmin}{argmin}

\usepackage{setspace,graphicx,epstopdf,amsmath,amsfonts,amssymb,amsthm,versionPO}
\usepackage{marginnote,datetime,enumitem,rotating,fancyvrb,scalerel}
\usepackage{hyperref}
\excludeversion{notes}		% Include notes?
\includeversion{links}          % Turn hyperlinks on?

\iflinks{}{\hypersetup{draft=true}}

\ifnotes{%
\usepackage[margin=1in,paperwidth=10in,right=2.5in]{geometry}%
\usepackage[textwidth=1.4in,shadow,colorinlistoftodos]{todonotes}%
}{%
\usepackage[margin=1in]{geometry}%
\usepackage[disable]{todonotes}%
}

\makeatletter\let\chapter\@undefined\makeatother % Undefine \chapter for todonotes

\usepackage{indentfirst} % Indent first sentence of a new section.
\usepackage{jfe}          % JFE-specific formatting of sections, etc.

\usepackage[
  backend=biber,
  style=numeric, % default
]{biblatex}
\usepackage{float}
\usepackage[caption = false]{subfig}

\begin{document}

\setlist{noitemsep}  % Reduce space between list items (itemize, enumerate, etc.)
%\onehalfspacing      % Use 1.5 spacing
% Use endnotes instead of footnotes - redefine \footnote command

\title{\textbf{Multitask Learning Deep Neural Networks to Combine Revealed and Stated Preference Data}}

\author{Shenhao Wang; Qingyi Wang; Jinhua Zhao \\
  Massachusetts Institute of Technology \\
  \\
  Aug 2019 \\
  }

\date{}              % No date for final submission

% Create title page with no page number
\renewcommand{\thefootnote}{\fnsymbol{footnote}}
\singlespacing
\maketitle

\vspace{-.2in}
\begin{abstract}
\noindent
It is an enduring question how to combine revealed preference (RP) and stated preference (SP) data to analyze travel behavior. This study presents a framework of multitask learning deep neural networks (MTLDNNs) for this question, and demonstrates that MTLDNNs are more generic than the traditional nested logit (NL) method, due to its capacity of automatic feature learning and soft constraints. About 1,500 MTLDNN models are designed and applied to the survey data that was collected in Singapore and focused on the RP of four current travel modes and the SP with autonomous vehicles (AV) as the one new travel mode in addition to those in RP. We found that MTLDNNs consistently outperform six benchmark models and particularly the classical NL models by about 5\% prediction accuracy in both RP and SP datasets. This performance improvement can be mainly attributed to the soft constraints specific to MTLDNNs, including its innovative architectural design and regularization methods, but not much to the generic capacity of automatic feature learning endowed by a standard feedforward DNN architecture. Besides prediction, MTLDNNs are also interpretable. The empirical results show that AV is mainly the substitute of driving and AV alternative-specific variables are more important than the socio-economic variables in determining AV adoption. Overall, this study introduces a new MTLDNN framework to combine RP and SP, and demonstrates its theoretical flexibility and empirical power for prediction and interpretation. Future studies can design new MTLDNN architectures to reflect the speciality of RP and SP and extend this work to other behavioral analysis.
\end{abstract}

\medskip

%\noindent \textit{JEL classification}: XXX, YYY.
%\medskip
%\textit{Keywords}: \LaTeX; papers with no content.

\thispagestyle{empty}

\clearpage

\onehalfspacing
\setcounter{footnote}{0}
\renewcommand{\thefootnote}{\arabic{footnote}}
\setcounter{page}{1}

\section{Introduction}
\label{s:1}
\noindent
Both revealed preference (RP) and stated preference (SP) data are widely used for demand analysis with their own pros and cons. RP data are commonly thought to have stronger external validity, but problematic owing to the limited coverage and high correlation of attributes. SP data are necessary when researchers seek to understand the effects of new attributes or alternatives, while they have biases owing to respondents' sensitivity to survey formats and unrealistic hypothetical scenarios. To mitigate these problems, researchers often combine them by using a nested logit (NL) approach, which assigns the alternatives in RP and SP to two nests with different scale factors \cite{Hensher1993,Bradley1997,Ben_Akiva1990,Ben_Akiva1994} \footnote{This nested logit method can also be seen as a pooled estimation with heteroscedasticity across RP and SP \cite{Louviere1999,Helveston2018}}. This NL approach is used to predict future travel demand and examine the factors that determine the adoption of certain travel alternatives based on parameter estimation. However, this NL method heavily relies on handcrafted feature engineering based on domain knowledge, which could be too restrictive in comparison to the automatic feature learning in deep neural networks (DNNs), as shown in many empirical studies \cite{LeCun2015,Bengio2013,Collobert2008}. This capacity of automatically learning features is enabled by the theoretically appealing property of DNN being a universal approximator \cite{Hornik1989,Hornik1991,Cybenko1989}, and as a result, DNN has demonstrated its extraordinary prediction power across the domains of natural language processing, image recognition, and travel behavioral analysis \cite{Fernandez2014,Krizhevsky2012,LeCun2015}. The theoretically appealing property and the empirically predictive power of DNN prompt us to ask whether it is possible to address the classical problem of combining RP and SP for demand analysis in a DNN framework, in a way more generic and flexible than the traditional NL method.

\begin{figure}[htb]
\centering
{\includegraphics[width=0.6\linewidth]{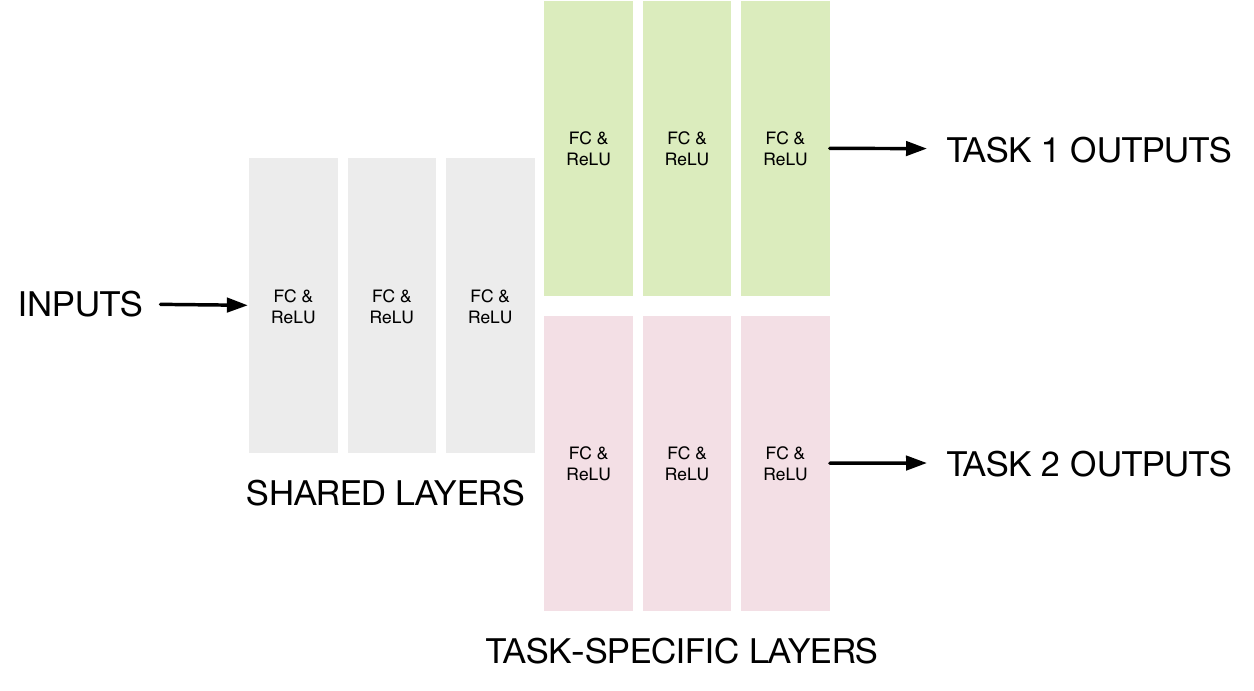}\label{fig:mtldnn}}\\
\caption{MTLDNN Architecture; this architecture has $3$ shared and $3$ task-specific layers, but they represent generally $M_1$ shared and $M_2$ task-specific layers; FC stands for fully connected layers; ReLU for Rectified Linear Units.}
\label{fig:mtldnn_arch}
\end{figure}

This paper presents the multitask learning deep neural network (MTLDNN) framework to jointly model RP and SP as two different but relevant tasks. One MTLDNN architecture is visualized in Figure \ref{fig:mtldnn_arch}, which starts with shared layers and ends with task-specific layers, capturing both the similarities and differences between tasks \cite{Caruana1997}. This architecture is more generic than the classical NL method, because it has the capacity of automatic feature learning and soft constraints. The automatic feature learning in MTLDNN referes to the process of automatically learning the feature transformation based on a powerful model class assumption (e.g. DNN), as opposed to the handcrafted feature engineering in the NL that relies on researchers' prior knowledge for model specification. The soft constraints refer to the flexibility of MTLDNN architectures and the regularization methods used in the training process of MTLDNNs, as opposed to the hard constraints such as parameter sharing between tasks, as commonly done in the NL approach. Specifically, the prototype MTLDNN architecture in Figure \ref{fig:mtldnn_arch} is flexible because it could take various forms with different shared and task-specific layers, which are designed into the hyperparameter space of the MTLDNN model. 

This MTLDNN framework is further examined in a dataset collected in Singapore, which studies the adoption of autonomous vehicles (AV) by adding the AV travel mode to four revealed travel modes in a SP survey. In this application, both the top 1 and top 10 ensembled MTLDNN models consistently outperform six other models, including NL with parameter constraints (NL-C), NL with no parameter constraints (NL-NC), separate multinomial logit models (MNL-SPT), joint multinomial logit models for RP and SP (MNL-JOINT), separate DNN models (DNN-SPT), and joint DNN models for RP and SP (DNN-JOINT). Specifically, MTLDNNs outperform the classical benchmark NL-NC and NL-C models by about $5 \%$ prediction accuracy in joint RP and SP, individual RP, and individual SP datasets. To attribute this performance to potential causes, we compare the eight models in a pairwise manner, removing the confounding differences between MTLDNN and NL methods. We found that this $ 5 \%$ prediction accuracy gain is mainly caused by the soft constraints designed in the MTLDNN framework, while the naive application of standard DNN architectures does not improve the model performance. To interpret the MTLDNN for AV adoption, we visualize the relationship between choice probability functions and input variables and computed the elasticities by using the gradient information of MTLDNNs, as commonly done in the machine learning community to reveal the information of DNN models \cite{Baehrens2010,Simonyan2013}. The results show how the travel modes substitute each other as alternative-specific and individual-specific variables change their values. This process of interpreting MTLDNNs demonstrates that MTLDNNs can not only predict more accurately, but also provide key insights at least at the same level as the classical tools.

As far as we know, this is the first study that presents the MTLDNN framework to combine RP and SP in demand analysis. Even with this simplest MTLDNN framework (Figure \ref{fig:mtldnn_arch}), our analysis has shown its power in terms of theoretical properties and empirical prediction. Since this MTLDNN architecture is the simplest among many possibilities, further studies could explore a variety of other MTLDNN architectures in the task of combining RP and SP, achieving better prediction and interpretation \cite{LongMingsheng2015,Hashimoto2016,Misra2016,Ruder2017_sluice}. The MTLDNN framework can also be used to jointly analyze car ownership and travel mode choice \cite{Train1980,Zegras2010}, activity patterns and trip chain choices \cite{Kitamura1992,Golob1997}, and many other applications that are traditionally analyzed by structural equation models (SEM). Moreover, the similarity of the visual structure between MTLDNNs and NL models prompts us to ask whether MTLDNNs can be seen as a generic extension of NL, which we cannot answer here but believe as an interesting question for future studies. 

The paper is organized as following. Section 2 reviews the MTLDNN models and the traditional methods of combining RP and SP. Section 3 presents the MTLDNN model and discusses why it is more generic than the NL method. Section 4 presents the setup of experiments, and Section 5 analyzes model performance, sources of the prediction gain of MTLDNNs, and the economic information in MTLDNNs. Section 6 concludes our findings and discuss future research directions.

\section{Literature Review}
\label{s:2}
\noindent
RP and SP data are both necessary for travel demand analysis, but they suffer from different sources of problems. The RP data is problematic due to its limited coverage of attributes, high correlation between attributes, and poor quality of background information \cite{Ben_Akiva1994}, although it clearly has a better external validity. As to the SP data, respondents could fail to provide valid answers because of respondents' sensitivity to survey formats, unrealistic hypothetical scenarios \cite{Small1998}, dynamics between RP and reported attitudes \cite{Small2007td}, or even just measurement errors that happen in the data collection process \cite{Hausman1998,Hausman2001}, although SP is the only possible data source for testing the effects of new pricing strategies, new public transit services, or new travel modes \cite{Ben_Akiva1990,Polydoropoulou_Ben_Akiva2001}. 

One common remedy is to jointly estimate RP and SP, with the benefits of gaining efficiency and correcting biases \cite{Ben_Akiva1994}. More specifically, researchers often use the NL model to combine RP and SP by treating RP and SP as two nests of choices \cite{Hensher1993,Bradley1997,Polydoropoulou_Ben_Akiva2001,Morikawa2002}. For instance, Polydoropoulou and Ben-Akiva (2001) used the NL approach to analyze the travel mode choice for multiple mass transit technologies. Golob et al. (1997) \cite{Golob1997_rp_sp} used the same method to examine how vehicle usage depends on the factors of vehicles and fuel types. In these NL modeling practices, researchers need to make parametric assumptions to capture the differences and similarities between RP and SP \cite{Bradley1997}. The first common assumption is about to what extent the RP and SP choice models share the same parameters; for instance, price and time coefficients in RP and SP could be assumed the same \cite{Polydoropoulou_Ben_Akiva2001}. The second common assumption is that RP and SP have different randomness in their error terms, causing the different magnitudes of the coefficients \cite{Hensher1993,Bradley1997}. While details of the modeling specifics may vary with studies, these assumptions along with the NL approach developed in the 1990s have become one standard way of combining RP and SP in choice modeling \cite{Small1998,Louviere1999,Whitehead2008,Train2009}. 

From a machine learning perspective, combining RP and SP could be addressed by using the MTLDNN framework because RP and SP can be treated as two different but highly related tasks. This MTLDNN framework has shown a great success when applied to many specific fields, including natural language processing (NLP) \cite{Collobert2008,Hashimoto2016}, image recognition \cite{LongMingsheng2015,Misra2016}, and healthcare for massive drug discovery \cite{Ramsundar2015}. As opposed to the models trained separately for each task, MTLDNNs improve generalization because it leverages the multiple sources of domain-specific information and introduces an inductive bias incorporated in the signals of various data sources \cite{Caruana1997}. In addition, ”multiple tasks arise naturally” in the real world \cite{Caruana1997}, so MTLDNN should be a natural choice when several tasks are closely or loosely related. 

Specifically, Caruana (1997) \cite{Caruana1997} first created a benchmark MTLDNN architecture, which starts with shared layers and ends with task-specific layers, as used in this study (Figure \ref{fig:mtldnn_arch}). This MTLDNN architecture has been applied to natural language processing tasks, showing the state-of-the-art performance in the absence of handcrafted feature engineering \cite{Collobert2008}. Caruana's initial MTLDNN architecture has been later improved by new MTLDNN architectural designs, such as deep relationship network (DRN) \cite{LongMingsheng2015}, hierarchical MTLDNN \cite{Hashimoto2016}, cross-stitch network \cite{Misra2016}, and SLUICE network \cite{Ruder2017_sluice}. Similar to the modeling concerns in the NL method, all the MTLDNN studies concern how to control the similarities and differences of the multiple tasks, and this concern can be addressed by designing specific architectures or regularizations. Caruana's MTLDNN framework addresses this concern by using the first several layers to reflect the similarities and the following task-specific layers to capture the differences \cite{Caruana1997}. Studies after Caruana's work explicitly designed MTLDNN architectures to capture both positive and negative relationship between tasks \cite{Misra2016,Ruder2017_sluice}. In addition to architectural design, researchers also used various regularizations to control the parameter distances between tasks, such as using $L_1/L_p$ group LASSO \cite{Argyriou2007,YuanMing2006}, mean and variance regularization \cite{Evgeniou2005,Jacob2009}, trace norm regularization \cite{YangYongxin2016}, tree-guided regularization \cite{Kim2010}, or Bayesian tensor normal priors \cite{LongMingsheng2015}. While sharing some similarity with the parameter constraints in the classical NL method, these constraints are much more generic and flexible in describing the relationship between tasks.

\section{Theory}
\noindent
This section introduces the MTLDNN model and the NL method, discusses why MTLDNNs are more generic than NL, and briefly discusses the potential weakness of MTLDNN by using statistical learning theory.

\subsection{Multitask Learning Deep Neural Network for RP and SP}
\noindent
Let $x_{r,i}$ and $x_{s,t}$ denote the input variables for RP and SP; $r$ and $s$ stand for RP and SP, $i \in \{ 1, 2, ... , N_r \}$ and $t \in \{ 1, 2, ... , N_s \}$ are the index of RP and SP observations. $x_{r,i}, x_{s,t} \in R^d$, in which $d$ represents the input dimension. The output choices of RP and SP are denoted by $y_{r,i}$ and $y_{s,t}$; $y_{r,i} \in \{0,1\}^{K_r}$ and $y_{s,t} \in \{0,1\}^{K_s}$; $K_r$ and $K_s$ are the dimension of outputs. SP often has more alternatives than RP since SP includes new products that are not available in the existing market ($K_s > K_r$). Both $y_{r,i}$ and $y_{s,t}$ are vectors taking zero or one values, and each component in $y_{r,i}$ and $y_{s,t}$ is denoted by $y_{k_r,i} \in \{0,1\}$ and $y_{k_s,t} \in \{0,1\}$. Due to the constraint of mutually exclusive and collectively exhaustive alternatives, $\sum_{k_s} y_{k_s,t} = 1$ and $\sum_{k_r} y_{k_r,i} = 1$. $k_r$ and $k_s$ are the index of alternatives in RP and SP, so $k_r \in \{1, 2, ..., K_r \}$ and $k_s \in \{1, 2, ..., K_s \}$. As represented by Figure \ref{fig:mtldnn_arch}, the feature transformation of RP and SP can be represented by the following:
\begin{flalign}
& V_{k_r, i} = (g_r^{M_2, k_r} \circ g_r^{M_2 - 1} \circ ... \circ g_r^{1}) \circ (g_0^{M_1} \circ g_0^{M_1 - 1} \circ ... \circ g_0^{1})(x_{r,i}) \label{eq:deterministic_util_mtldnn_rp} \\
& V_{k_s, t} = (g_s^{M_2, k_s} \circ g_s^{M_2 - 1} \circ ... \circ g_s^{1}) \circ (g_0^{M_1} \circ g_0^{M_1 - 1} \circ ... \circ g_0^{1})(x_{s,t}) \label{eq:deterministic_util_mtldnn_sp}
\end{flalign}

\noindent
in which $M_1$ represents the depth of the shared layers and $M_2$ the depth of the task-specific layers; $g_0$ represents the transformation of one shared layer; $g_r$ and $g_s$ represent the transformation of one layer in RP and SP. Specifically, $g$ functions (including $g_r$, $g_s$, and $g_0$) are the composition of ReLU and linear transformation: $g^l(x) = \max \{W^l x, 0 \}, \ \forall l \neq M_2$. Equations \ref{eq:deterministic_util_mtldnn_rp} and \ref{eq:deterministic_util_mtldnn_sp} describe precisely the MTLDNN architecture in Figure \ref{fig:mtldnn_arch}: $(g_0^{M_1} \circ g_0^{M_1 - 1} \circ ... \circ g_0^{1})$ represent the shared layers, while $(g_r^{M_2, k_r} \circ g_r^{M_2 - 1} \circ ... \circ g_r^{1})$ and $(g_s^{M_2, k_s} \circ g_s^{M_2 - 1} \circ ... \circ g_s^{1})$ represent task-specific layers. As a result, the choice probability functions in RP and SP can be represented by
\begin{flalign}
& P(y_{k_r, i}; w_r, w_0) = \frac{e^{V_{k_r, i}}}{\sum_{j_r = 1}^{K_r} e^{V_{j_r, i}}} \label{eq:choice_probability_mtldnn_rp} \\
& P(y_{k_s, t}; w_s, w_0, T) = \frac{e^{V_{k_s, t}/T}}{\sum_{j_s = 1}^{K_s} e^{V_{j_s, t}/T}} \label{eq:choice_probability_mtldnn_sp}
\end{flalign}

\noindent
in which $w_r$ and $w_s$ represent the task-specific parameters in $g_r$ and $g_s$; $w_0$ the shared parameters in $g_0$. Equation \ref{eq:choice_probability_mtldnn_rp} takes the form of a standard Softmax activation function, while that of SP (Equation \ref{eq:choice_probability_mtldnn_sp}) is adjusted by the $T$ factor, which is referred to as ``temperature'' in the DNN literature to change the scale of logits \cite{Hinton2015}. 

With choice probabilities formulated, we train the model by minimizing the empirical risk (ERM):
\begin{equation}
\begin{aligned}
\underset{w_r, w_s, w_0, T}{\min} R(X,Y; w_r, w_s, w_0, T; c_H) = \underset{w_r, w_s, w_0, T}{\min} 
	& \Big\{ - \frac{1}{N_r} \sum_{i = 1}^{N_r} \sum_{k_r = 1}^{K_r} y_{k_r} \log P(y_{k_r, i}; w_r, w_0; c_H) \\
	& \ \ \ \ - \frac{\lambda_0}{N_s} \sum_{t = 1}^{N_s} \sum_{k_s = 1}^{K_s} y_{k_s} \log P(y_{k_s, t}; w_r, w_0, T; c_H) \\
	& \ \ \ \ + \lambda_1 ||w_0||^2_2 + \lambda_2 ||w_s||^2_2 + \lambda_3||\tilde{w}_s - w_r||_2^2 \Big\}
\end{aligned}
\label{eq:erm_mtldnn}
\end{equation}

\noindent
Equation \ref{eq:erm_mtldnn} consists of three parts: the first part $- \frac{1}{N_r} \sum_{i = 1}^{N_r} \sum_{k_r = 1}^{K_r} y_{k_r} \log P(y_{k_r, i}; w_r, w_0; c_H)$ is the empirical risk of RP; the second part $- \frac{\lambda_0}{N_s} \sum_{t = 1}^{N_s} \sum_{k_s = 1}^{K_s} y_{k_s} \log P(y_{k_s, t}; w_r, w_0, T; c_H)$ is the empirical risk of SP; the third part $\lambda_1 ||w_0||^2_2 + \lambda_2 ||w_s||^2_2 + \lambda_3||\tilde{w}_s - w_r||_2^2$ is the explicit regularization. In total, Equation \ref{eq:erm_mtldnn} incorporates four hyperparameters ($\lambda_0$, $\lambda_1$, $\lambda_2$, $\lambda_3$) for explicit regularizations. $\lambda_0$ adjusts the ratio of empirical risks between RP and SP. This study treats equally one observation in RP and SP by fixing $\lambda_0 = 1$ \footnote{Researchers are free to choose the value of $\lambda_0$, since there is no clear-cut rule for its value specification. Our choice reflects our belief that each individual counts as equal in RP and SP.}. $\lambda_1$ and $\lambda_2$ jointly adjust the absolute magnitudes of the shared layers and SP-specific layers: larger $\lambda_1$ and $\lambda_2$ lead to larger weight decay, reducing the estimation error in the complex DNN models \cite{Vapnik1999}. $\lambda_3$ controls the degree of similarity between RP- and SP-specific layers. As $\lambda_3$ becomes very large, ERM penalizes more the large differences between RP- and SP-specific layers, leading to more similarities shared by the coefficients in RP and SP models. Since $w_s$ and $w_r$ do not match perfectly in our case, $\tilde{w}_s$ is used to denote the SP-specific weights that are corresponding to those RP-specific weights. This specific ERM formulation and the regularizations in Equation \ref{eq:erm_mtldnn} are commonly used in MTLDNN studies \cite{Evgeniou2005,Jacob2009}. 

\subsection{Nested Logit Model for RP and SP}
\noindent
The NL method comes from the past studies \cite{Hensher1993,Bradley1997,Polydoropoulou_Ben_Akiva2001,Morikawa2002}. The utility functions in RP and SP are assumed to be
\begin{flalign}
U_{k_r, i} = V_{k_r, i} + \epsilon_{k_r} = \beta^T_{k_r} \phi(x_{r, i}) + \epsilon_{k_r, i} \label{eq:rp_util_nested_logit}\\
U_{k_s, t} = V_{k_s, t} + \epsilon_{k_s} = \beta^T_{k_s} \phi(x_{s, t}) + \epsilon_{k_s, t} \label{eq:sp_util_nested_logit}
\end{flalign}

\noindent
in which $\beta_{k_r}$ and $\beta_{k_s}$ are the parameters for RP and SP; $\phi$ denotes the feature transformation based on domain-specific knowledge; for example, $\phi$ can represent the quadratic transformation, when researchers believe there exists nonlinear relationship between utilities and input variables. $\epsilon_{k_r, i}$ and $\epsilon_{k_s, t}$ are random utility terms. It is commonly assumed that $\epsilon_{k_r, i}$ and $\epsilon_{k_s, t}$ are off by one scale factor:
\begin{flalign}
Var(\epsilon_{k_r, i})/Var(\epsilon_{k_s, t}) = 1/\theta^2
\label{eq:proportional_error_term}
\end{flalign}

\noindent
The choice probability functions based on the NL approach are:
\begin{flalign}
& P(y_{k_r, i}; \beta_r) = \frac{e^{\beta_{k_r}^T \phi(x_{r,i})}}{\sum_{j_r = 1}^{K_r} e^{\beta_{j_r}^T \phi(x_{r,i})}} \label{eq:choice_probability_nl_rp} \\
& P(y_{k_s, t}; \beta_s) = \frac{e^{\beta_{k_s}^T \phi(x_{s,t})/\theta}}{\sum_{j_s = 1}^{K_s} e^{\beta_{j_s}^T \phi(x_{s,t})/\theta}} \label{eq:choice_probability_mtldnn_nl_sp}
\end{flalign}

\noindent
Here $\beta_r$ and $\beta_s$ represent all the parameters in RP and SP. Note that $\theta$ is similar to the temperature factor $T$ in the MTLDNN framework, although $\theta$ here arises from the assumption about variance of the random error terms while $T$ does not. As a result, the ERM in NL is
\begin{flalign}
\underset{\beta_r, \beta_s}{\min} \ R(X,Y;\beta_r, \beta_s) = \underset{\beta_r, \beta_s}{\min} \Big\{ - \frac{1}{N} \big[ \sum_{i=1}^{N_r} \sum_{k_r = 1}^{K_r} y_{k_r, i} \log P(y_{k_r, i}; \beta_r) + \sum_{t=1}^{N_s} \sum_{k_s = 1}^{K_s} y_{k_s, t} \log P(y_{k_s, t}; \beta_s) \big] \Big\}
\label{eq:erm_nl}
\end{flalign}

This NL formulation is not the same as a standard NL model, since respondents do not face all the RP and SP alternatives in one choice scenario. Therefore, researchers named this NL approach as ``artificial nested logit'' model, the details of which are available in \cite{Hensher1993,Bradley1997}.

\subsection{MTLDNNs are More Generic Than NL}
\noindent
The MTLDNN framework is more generic than NL due to the capacity of automatic feature learning and the soft constraints, including the architectural design and regularization hyperparameters.

\begin{figure}[htb]
\centering
\subfloat[NL with $\phi(x)$]{\includegraphics[width=0.35\linewidth]{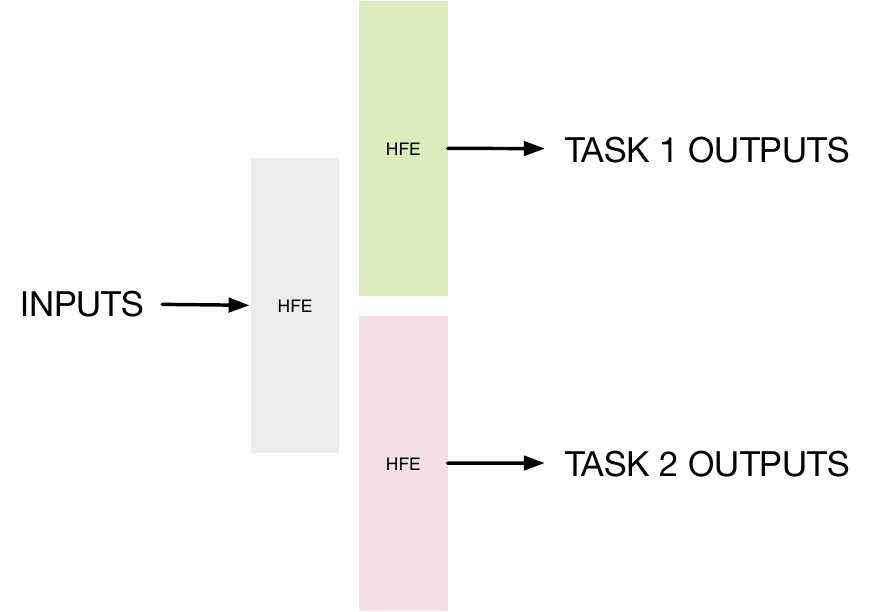}\label{sfig:nest1}}
\subfloat[NL without $\phi(x)$]{\includegraphics[width=0.3\linewidth]{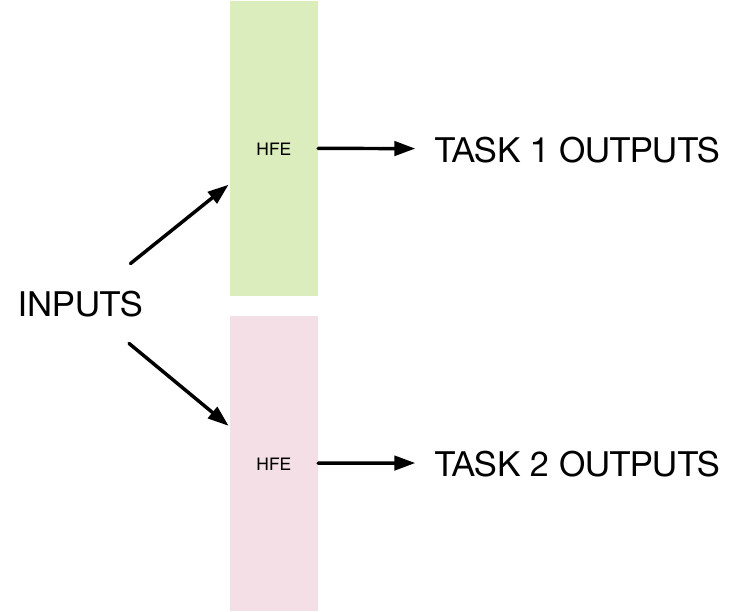}\label{sfig:nest2}}
\caption{Visualization of NL; HFE stands for handcrafted feature engineering}
\label{fig:nl_arch}
\end{figure}

First, the utility specification of MTLDNNs (Equations \ref{eq:deterministic_util_mtldnn_rp} and \ref{eq:deterministic_util_mtldnn_sp}) takes a layer-by-layer function form, which makes DNN a universal approximator and enables the capacity of automatic feature learning. This function form is in sharp contrast to NL, which strongly relies on the handcrafted feature mapping function $\phi()$ and linear-in-parameter specification $\beta_r$ and $\beta_s$ (Equations \ref{eq:rp_util_nested_logit} and \ref{eq:sp_util_nested_logit}). The handcrafted feature engineering is problematic because modelers' prior knowledge is rarely complete for the task at hand, and this incompleteness of knowledge leads to function misspecification error and low prediction accuracy in the NL approach. On the other side, the strong approximation power of MTLDNNs enables it to approximate any underlying behavioral mechanism without depending on the completeness of modelers' domain knowledge. In fact, Equations \ref{eq:rp_util_nested_logit} and \ref{eq:sp_util_nested_logit} can be visualized in Figure \ref{sfig:nest1}, in which the grey layer represents $\phi()$ transformation and the green and red layers represent $\beta_r$ and $\beta_s$ multiplication. When researchers only use the identity mapping for $\phi$ ($\phi(x) = x$), Equations \ref{eq:rp_util_nested_logit} and \ref{eq:sp_util_nested_logit} can be visualized in Figure \ref{sfig:nest2}, in which inputs are directly feed into task-specific layers. Therefore, other than the difference in terms of automatic vs. handcrafted feature learning, the difference between MTLDNNs and NL also reflects the difference between deep and shallow neural network (DNN vs. SNN). Studies show that DNN is a more efficient universal approximator than SNN \cite{Cohen2016,Liang2016,Rolnick2017}, and the benefit of depth could explain why MTLDNNs have stronger approximation power than NL.

Second, while both MTLDNNs and NL incorporate mechanisms to capture the similarities and differences between RP and SP, MTLDNNs have softer constraints than NL models. The soft constraints refer to two parts: the architectural design and regularization methods. In terms of architectural design, Figure \ref{fig:mtldnn_arch} is a prototype of various MTLDNN architectures, which vary in terms of the numbers of shared layers ($M_1$) and the numbers of task-specific layers ($M_2$). As a result, the diversity of architectural design empowers MTLDNN to learn underlying features in a way more flexible than NL, which only has a fixed shallow architecture, as shown in Figure \ref{sfig:nest1}. The regularization methods in MTLDNNs are also more flexible than NL. For example, in the NL model, researchers need to specify to what extent $\beta_{k_r}$ and $\beta_{k_s}$ are the same. Let $\beta_{k_r, j}$ and $\beta_{k_s, j}$ denote the coefficient for the price variable in RP and SP, researchers choose to impose no constraint $\beta_{k_r, j} \neq \beta_{k_s, j}$ or impose a hard constraint $\beta_{k_r, j} = \beta_{k_s, j}$ based on their prior belief. In other words, researchers must choose whether the coefficients of price estimated from the RP data should be the same as that from the SP data in a a-priori manner. This hard constraint can be equally represented by adding to Equation \ref{eq:erm_nl} a penalty term $|| \beta_{k_r, j} - \beta_{k_s, j} ||_2$ multiplied by a large $\lambda$, and the no constraint case is associated with $\lambda = 0$. Therefore, $\lambda_3$ term in Equation \ref{eq:erm_mtldnn} is a soft control on the similarity between RP and SP, and it incorporates the hard constraints in the NL as its two boundary points. As $\lambda_3$ ranges from zero to a large value, RP and SP models ranges from sharing no similarity to full similarity. The second example is the scale constraint in MTLDNNs vs. NL. While the $T$ factor is nearly the same as the $\theta$ factor in NL, it is important to note that the temperature factor is not the only hyperparameter that controls the difference of utility scale between RP and SP in the MTLDNN framework. Besides $T$, hyperparameteres $\lambda_1$ and $\lambda_2$ implicitly control the utility scale as well. When the overall magnitude of the parameters in MTLDNNs becomes larger (small $\lambda_1$ and $\lambda_2$), it is more likely for RP and SP to have larger utility scale difference, implying that the RP and SP have different randomness in their systems. 

\subsection{Approximation and Estimation Error of MTLDNNs}
\label{section:approx_est_mtldnn}
\noindent
MTLDNNs as a more generic model family does not necessarily imply a higher prediction accuracy, since the small approximation error obtained by larger model complexity may be counteracted by the large estimation error. Based on statistical learning theory, a more complex model typically has smaller approximation errors (bias) but larger estimation error (variance) \cite{Von_Luxburg2011,Vapnik1999} \footnote{This tradeoff is traditionally known as bias-variance tradeoff. Bias is similar to the approximation error, and variance is similar to the estimation error}. This problem could be one potential weakness of MTLDNNs, due to its much larger function complexity, formally measured by Vapnik-Chervonenkis (VC) dimension \cite{Vapnik2013,Vapnik1999}. Specifically, the VC dimension of DNNs is roughly proportional to its number of parameters and its depth, so the VC dimension of a simple 5-layer DNN with 100 neurons in each layer is $c_0 \times 250,000$ ($O(100^2 \times 5 \times 5)$) \cite{Bartlett2017}. On the other side, the VC dimension of a NL model is proportional to its number of parameters: with about $20$ input variables, the VC dimension of NL is only about $c_1 \times 20$. While this VC dimension perspective is not the optimum upper bound of the estimation error \cite{Golowich2017,Neyshabur2015}, it provides sufficient insights for the purpose of our paper \footnote{For a more general introduction, readers could refer to the recent studies in the fields of high dimensional probability and statistics \cite{Wainwright2019,Vershynin2018,Bartlett2002,Anthony2009}}. Generally speaking, while DNNs are more generic than multinomial logit models (MNL) in terms of the function class relationship \cite{Cohen2016,Liang2016,Rolnick2017}, DNNs could perform worse due to its high model complexity and its corresponding large estimation errors. Hence we need to conduct empirical experiments to evaluate the performance of MTLDNNs and NL.

\section{Experiment Setup}
\subsection{Data}
\noindent
An online survey was designed to explore the travel demand of AV and the underlying factors that determine the AV adoption. The online survey was collected from Singapore with the help of a professional survey company Qualtrics.com. The survey consisted of one section of revealed preference (RP) survey, one section of stated preference (SP) survey, and one section for eliciting socioeconomic variables. The travel mode alternatives in RP include walking, public transit, driving, and ride sharing; on-demand AV use was added to the SP survey as the additional travel mode. In total, we gathered 1,592 RP choice answers and 8,418 SP ones. 

\subsection{Training}
\noindent
RP and SP data are split into training and testing sets with the ratio of 4:1. One challenge in the MTLDNN training is its vast number of hyperparameters, and the performance of MTLDNN largely depends on hyperparameters. To address this problem, we specify a hyperparameter space and search randomly within this space to identify the hyperparameters that cause high prediction accuracy \cite{Bergstra2012}. Let $S_H$ denote the hyperparameter space. We sample one group of hyperparameters $c_H^{(q)}$ from $S_H$, and choose the one with the highest prediction accuracy in the testing set. Formally, 
\begin{equation}
\setlength{\jot}{2pt} \label{eq:hyper_searching}
  \begin{aligned}
  \hat{c}_H = \underset{c_H \in \{c_H^{(1)},c_H^{(2)}, ..., c_H^{(S)} \} }{\argmin} R(X,Y; \hat{w}_r, \hat{w}_s, \hat{w}_0, \hat{T}; c_H)
  \end{aligned}
\end{equation}

\noindent
in which $R(X,Y; \hat{w}_r, \hat{w}_s, \hat{w}_0, \hat{T}; c_H)$ is the estimated empirical risk (Equation \ref{eq:erm_mtldnn}); $S = 1,500$ represents the total number of random sampling in this study. Details of the hyperparameter space are incorporated in Appendix I; some descriptive summary statistics in Appendix II.

\section{Experiment Results}
\label{s:5}
\subsection{Model Performance}
\noindent
Table \ref{table:model_performance} summarizes the model performance of MTLDNN (Top 1), MTLDNN ensemble over top 10 models (MTLDNN-E), separate deep neural networks for RP and SP (DNN-SPT), deep neural networks for joint RP and SP (DNN-JOINT), NL with parameter constraints (NL-C), NL with no parameter constraints (NL-NC), separate multinomial logit models for RP and SP (MNL-SPT), and joint multinomial logit model for RP and SP (MNL-JOINT). In Table \ref{table:model_performance}, Panel 1 reports the joint prediction accuracy for RP and SP, individual RP, and individual SP data in the testing and training sets; Panel 2 summarizes the differences between the eight models in terms of four characteristics: automatic feature learning, soft constraints, hard constraints, and data augmentation. The models in the DNN family have the capacity of automatic feature learning; only two MTLDNNs have the soft constraints; only the NL with parameter constraints has the hard constraints; and the jointly trained models augment data and tasks. Overall, the six non-MTLDNN models are designed as benchmarks for performance comparison and for disentangling the reasons why MTLDNNs perform well.

\begin{table}[htb]
    \centering
    \resizebox{1.0\linewidth}{!}{%
    \begin{tabular}{@{} p{0.3\linewidth} P{0.1\linewidth} P{0.1\linewidth} P{0.1\linewidth} P{0.1\linewidth} P{0.1\linewidth} P{0.1\linewidth} P{0.1\linewidth} P{0.1\linewidth}@{}}
        \toprule
        \midrule
        \textbf{ } & \textbf{MTLDNN (Top1)} & \textbf{MTLDNN-E (Top10)} & \textbf{DNN-SPT} & \textbf{DNN-JOINT} & \textbf{NL-C} & \textbf{NL-NC} & \textbf{MNL-SPT} & \textbf{MNL-JOINT} \\
        \midrule
        \multicolumn{9}{c}{\small{Panel 1: Prediction Accuracy}} \\     
        \midrule
        Joint RP+SP (Testing) & 60.0\% & 58.7\% & 53.4\% & 53.8\% & 55.4\% & 55.0\% & 55.0\% & 51.9\% \\
        RP (Testing) 		  & 69.9\% & 66.6\% & 65.8\% & 65.8\% & 65.4\% & 64.7\% & 64.5\% & 44.0\% \\
        SP (Testing) 		  & 58.2\% & 57.2\% & 51.1\% & 51.5\% & 53.5\% & 53.2\% & 53.2\% & 53.5\% \\
        Joint RP+SP (Training) & 60.7\% & 62.2\% & 52.5\% & 52.9\% & 54.0\% & 54.5\% & 54.4\% & 50.3\% \\
        RP (Training) 		  & 69.1\% & 71.9\% & 59.8\% & 59.8\% & 58.9\% & 62.2\% & 62.1\% & 37.0\% \\
        SP (Training) 		  & 59.1\% & 60.3\% & 51.1\% & 51.5\% & 53.0\% & 53.0\% & 53.0\% & 52.8\% \\
        \midrule
        \multicolumn{9}{c}{\small{Panel 2: Different Characteristics of Models}} \\         
        \midrule
        Automatic Feature Learning & $\times$ & $\times$ & $\times$ & $\times$ & & & & \\ 
        Soft Constraints & $\times$ & $\times$ & & & & & & \\ 
		Hard Constraints &  & & &  & $\times$ & & & \\ 
		Data Augmentation & $\times$ & $\times$ & & $\times$ & $\times$ & $\times$ & & $\times$ \\
        \midrule
%        \multicolumn{9}{c}{\small{Notes: $\times$ implies the existence of the corresponding characteristic.}}
        \bottomrule
    \end{tabular}
    }
\caption{Comparison of Eight Models}
\label{table:model_performance}
\end{table}

Two MTLDNNs perform better than all the other six models in terms of joint and separate RP and SP prediction accuracy, as shown in Panel 1 in Table \ref{table:model_performance}. In terms of the joint prediction accuracy, the top $1$ MTLDNN model outperforms the NL models with and without parameter constraints by $4.5 \%$ and $5.0 \%$ in the testing set of the joint RP and SP data. This about $5 \%$ prediction gain of MTLDNNs over NL models is consistent in the out-of-sample performance of the separate RP and SP datasets. The top 10 MTLDNN model ensemble also has higher prediction accuracy than all the other models in the testing set of joint RP and SP and separate RP and SP datasets, although MTLDNN-E performs about $1.3 \%$ worse than the top 1 MTLDNN model. Note that the MTLDNN models not only outperform classical choice modeling methods such as NL and MNL, they also perform better than the DNN models without the MTLDNN architectures, such as DNN-SPT and DNN-JOINT, demonstrating the importance of the soft constraints in MTLDNN models. While the performance improvement of MTLDNNs over NL models is clear, the next question is how to attribute this improvement to two potential factors: automatic feature learning and soft constraints.

\subsection{Sources of Performance Improvement in MTLDNNs}
\noindent
It is difficult to disentangle the two factors by directly comparing MTLDNNs and two NL models since they are different by both, and even more difficult, NL-C has the hard constraint while MTLDNNs don't have, as shown in Panel 2 in Table \ref{table:model_performance}. To disentangle the effect of these factors, we compare the eight models in a pair-wise manner, making each pair differ by only one factor. In fact, the models in Table \ref{table:model_performance} have been roughly sorted with decreasing model complexity, from MTLDNNs, to DNNs, to NL, and lastly to MNL models. The comparison between models with similar model structures enables us to see the impact of individual factors.

\textbf{Automatic Feature Learning.} Comparing models between MNL-SPT and DNN-SPT, and between MNL-JOINT and DNN-JOINT helps to understand how automatic feature learning contributes to the performance improvement. Interestingly, DNN-SPT and DNN-JOINT do not outperform MNL-SPT and MNL-JOINT, suggesting that the straightforward application of the feedforward DNN architecture does \textit{not} improve the performance in MTLDNNs. Specifically, DNN-SPT performs worse than MNL-SPT by 1.6\% prediction accuracy and DNN-JOINT performs better than MNL-JOINT by 1.9\%. By applying the perspective of statistical learning theory (Section \ref{section:approx_est_mtldnn}), this result suggests that the large estimation error loss exceeds the gain of prediction accuracy on the approximation error side, at least in this one specific dataset. This result could happen when the underlying data generating process (DGP) is similar to that of MNL, leading to only trivial or zero reduction of approximation error by using DNN to replace MNL. While this result is different from many studies that found DNN outperforming MNL in the travel behavioral analysis \cite{Nijkamp1996,XieChi2003}, this type of finding is not unseen in previous studies \cite{Mozolin2000}. However, from another perspective, the performance of DNN implies that it cannot help the model performance if researchers only naively apply the default feedforward DNN architecture and hope to solve domain-specific problems without any adjustment. In fact, many of the recent new models in the deep learning community come from the innovation of DNN architectures \cite{Krizhevsky2012,Szegedy2015,HeKaiming2016}, and this MTLDNN case is no exception. 

\begin{figure}[htb]
\centering
\subfloat[Six Different Architectures: (5-0);(4-1);(3-2);(2-3);(1-4);(0-5)]{\includegraphics[width=1.0\linewidth]{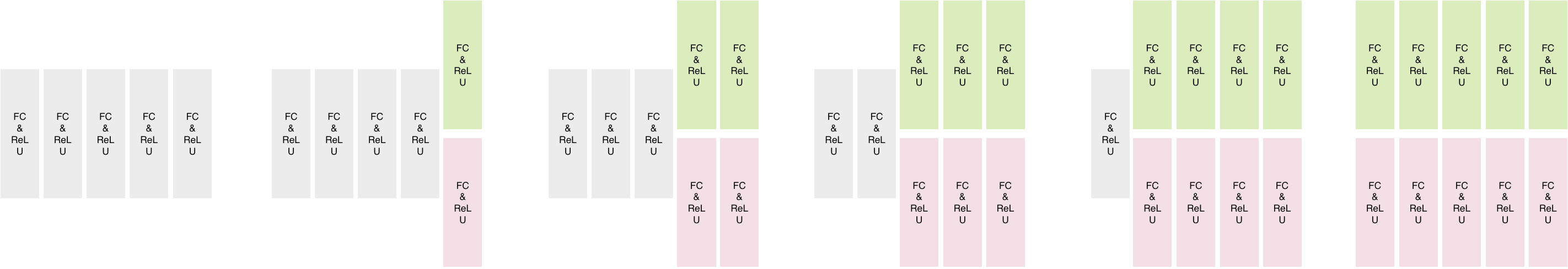}\label{sfig:mtldnn_shared_specific}} \\
\subfloat[Performance of Six Architectures]{\includegraphics[width = 0.4\linewidth]{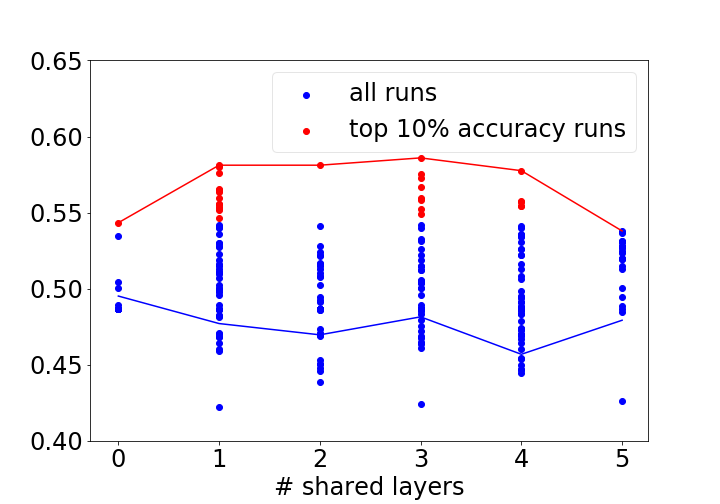}\label{sfig:mtldnn_shared_specific_performance}}
\subfloat[Prediction Accuracy and $\lambda_3$]{\includegraphics[width = 0.4\linewidth]{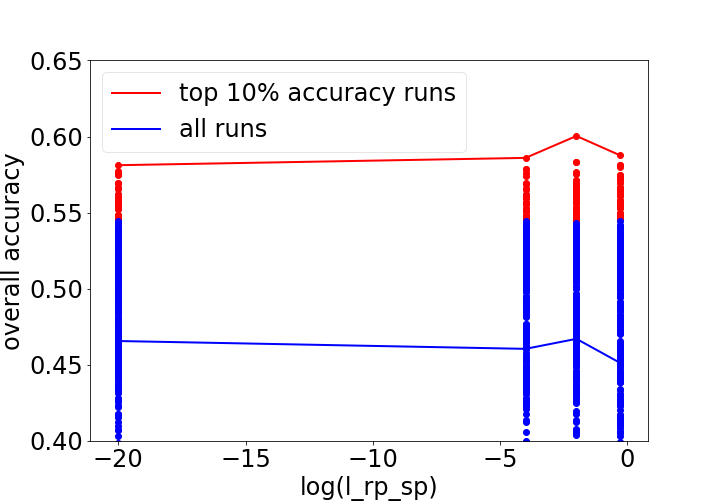}\label{sfig:mtldnn_lambda_3}} \\
\caption{Architectural Design and Regularization; red and blue dots are the results of individual models; blue lines connect average prediction accuracy of all models, and red ones connect the models with the maximum prediction accuracy.}
\label{fig:mtldnn_s_s}
\end{figure}

\textbf{Soft Constraints.} Comparing MTLDNNs and DNNs enables us to understand the importance of soft constraints in MTLDNNs, particularly the MTLDNN-specific architectures and regularization methods. In fact, DNN-SPT and DNN-JOINT can be seen as two special cases of MTLDNN models. Figure \ref{sfig:mtldnn_shared_specific} shows a spectrum of MTLDNN examples, indexed by their shared vs. task-specific layers. On the left hand of Figure \ref{sfig:mtldnn_shared_specific}, MTLDNNs become DNN-JOINT, which has only five shared layers without any task-specific layer. On the right hand of Figure \ref{sfig:mtldnn_shared_specific}, MTLDNNs become DNN-SPT, which has only five task-specific layers without any shared layer. Between DNN-JOINT and DNN-SPT, MTLDNNs take various forms of architectures, varying with the ratio of shared vs. task-specific layers. Figure \ref{sfig:mtldnn_shared_specific_performance} visualizes the prediction accuracy of the six MTLDNN architectures in Figure \ref{sfig:mtldnn_shared_specific}. In this set of MTLDNN models with $5$ layers in total, the MTLDNN models with non-zero shared or task-specific layers perform substantially better than DNN-SPT and DNN-JOINT, and particularly the MTLDNN architecture with $3$ shared layers and $2$ task-specific layers performs the best. This result is the same as Table \ref{table:model_performance}, which shows that top 1 MTLDNN outperforms DNN-SPT and DNN-JOINT by $6.6 \%$ and $6.2 \%$ (average about $6.4 \%$) in the joint performance of RP and SP. These results demonstrate the effectiveness of MTLDNN-specific architectures (Figure \ref{fig:mtldnn_arch}), in contrast to the ineffectiveness of standard feedforward DNN architecture. In addition to architectural design, the regularization methods designed specifically for MTL are also helpful. Figure \ref{sfig:mtldnn_lambda_3} shows how prediction accuracy depends on $\lambda_3$, which is the penalty term on the similarity between the task-specific layers of RP and SP. When $\lambda_3$ becomes too large or small, which implies that task-specific layers between RP and SP are either too similar or different, the MTLDNN model cannot perform well. Only when RP and SP parameters differ to an appropriate degree, with $\lambda_3$ taking the value $0.01$ in our example, the MTLDNN architecture could achieve the optimum performance. Overall, these MTLDNN-specific soft constraints, both architectural design and regularization methods, have dramatically contributed to the performance improvement of MTLDNNs.

\textbf{Hard Constraints.} As a comparison to soft constraints, we examine whether hard constraints in NL-C help to improve the model performance of NL-NC, since hard constraints are most commonly used in classical methods of combining datasets from different sources. As shown in Table \ref{table:model_performance}, the model performances of NL-C and NL-NC are quite similar in that the joint performance of RP and SP in NL-C and NL-NC is only different by $0.4 \%$, which is much smaller than the $5 \%$ accuracy difference between MTLDNN and NL models and the about $6.4 \%$ accuracy difference between MTLDNNs and DNNs. This result is consistent with our previous discussions: hard constraints imposed by domain experts are often limited, and they are less generic and flexible than the soft constraints in the MTLDNN models. 

\textbf{Data Augmentation.} Data augmentation is sometimes believed to be one cause of MTLDNNs' high performance \cite{Caruana1997}, especially when researchers focus on the performance of one task rather than the joint performance. It is because MTLDNNs essentially augment more observations and tasks for predicting one task, as opposed to a model with only observations collected for this one task. \footnote{This data augmentation idea can be reframed as increasing estimation efficiency in the classical statistical discussions \cite{Ben_Akiva1994}.} Comparing DNN-SPT and DNN-JOINT or MNL-SPT and MNL-JOINT helps to identify the effectiveness of data augmentation. It turns out that data augmentation itself appears to show no effects on the performance improvement. Comparing DNN-SPT and DNN-JOINT, DNN-JOINT does not improve the performance of DNN-SPT in separate RP and SP datasets, although DNN-JOINT does use a larger sample size than DNN-SPT for either RP and SP tasks. The result is also similar for MNL-SPT and MNL-JOINT. The cause of this result, we believe, is again about the constraints between tasks. While joint training literally uses more observations than separate training, inappropriate constraints added to two tasks could distort the joint model and worsen the performance. 

\begin{figure}[htb]
\centering
\subfloat[Temperature Distribution]{\includegraphics[width=0.3\linewidth]{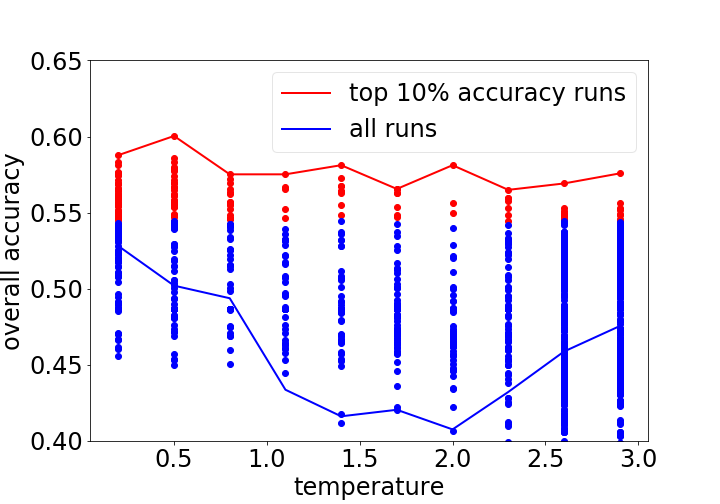}\label{sfig:hyper_temperature}}
\subfloat[$\lambda_1$]{\includegraphics[width=0.3\linewidth]{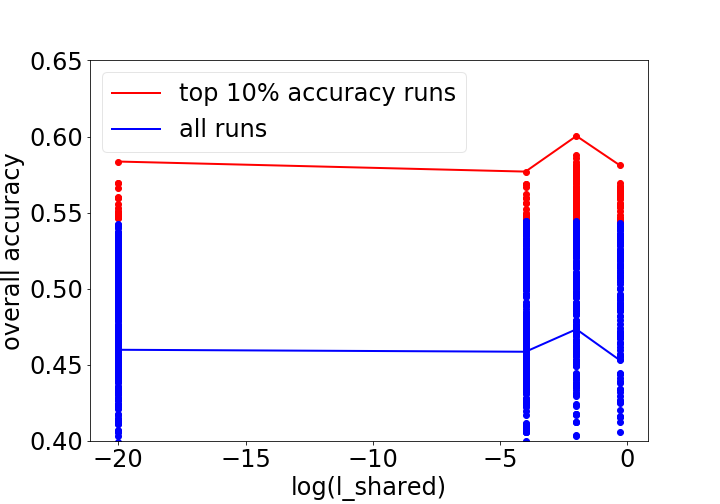}\label{sfig:hyper_lambda_1}} 
\subfloat[$\lambda_2$]{\includegraphics[width=0.3\linewidth]{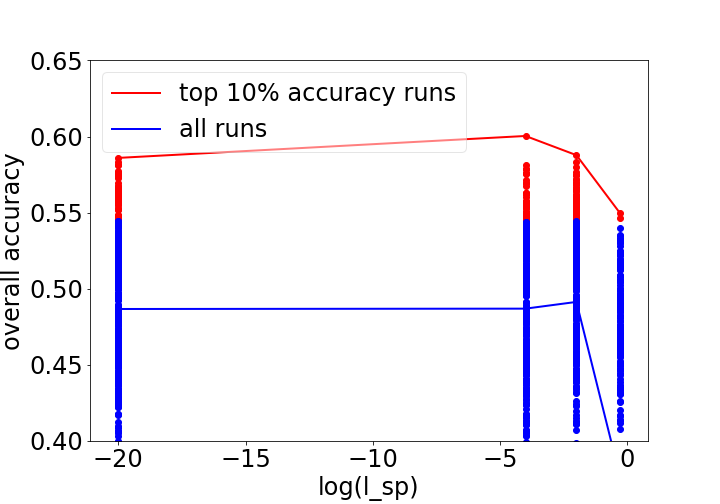}\label{sfig:hyper_lambda_2}} \\
\subfloat[Depth (Shared)]{\includegraphics[width=0.3\linewidth]{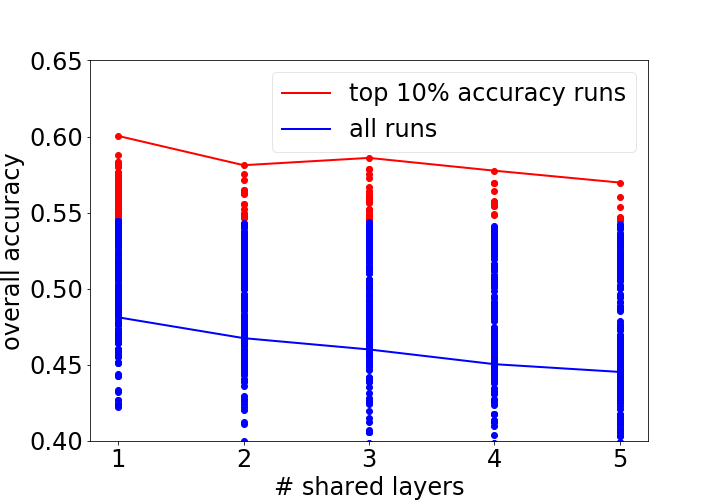}\label{sfig:hyper_depth_shared}}
\subfloat[Depth (Specific)]{\includegraphics[width=0.3\linewidth]{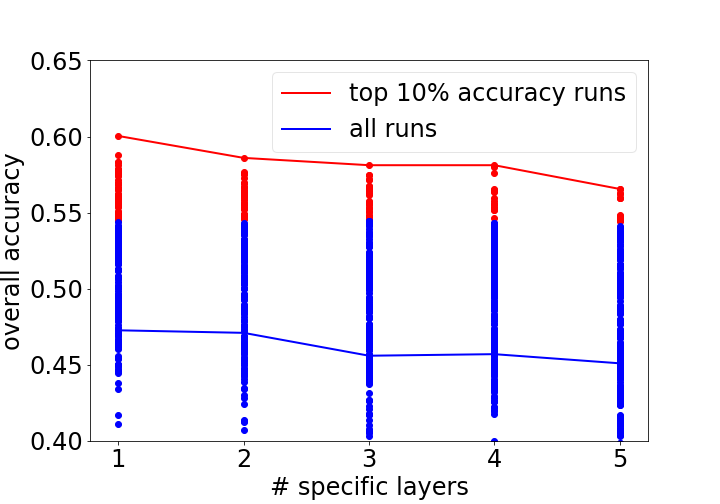}\label{sfig:hyper_depth_specific}}
\subfloat[Width]{\includegraphics[width=0.3\linewidth]{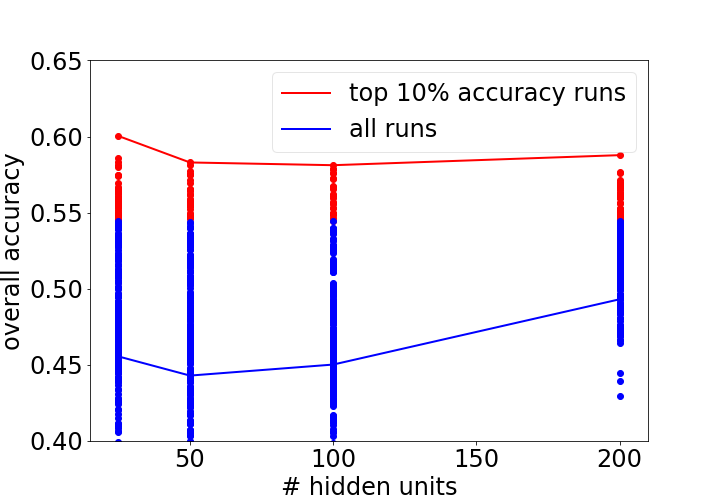}\label{sfig:hyper_width}} \\
\caption{Other Hyperparameters; red and blue dots are the results of individual models; blue lines connect average prediction accuracy of all models, and red ones connect the models with the maximum prediction accuracy.}
\label{fig:other_hyper}
\end{figure}

\textbf{Other Hyperparameters.} Figure \ref{fig:other_hyper} shows the role of other hyperparameters of MTLDNN in influencing its prediction accuracy \footnote{The hyperparameters of the top $10$ MTLDNN models are incorporated in Appendix III}. (1) Temperature (Figure \ref{sfig:hyper_temperature}). While we design the temperature factor into the MTLDNN framework by emulating the scale factor in NL models, the results show that the temperature factor does not matter than much. The top $10 \%$ models could have temperature values ranging from $0.2$ to $3.0$. (2) $\lambda_1$ and $\lambda_2$ (Figures \ref{sfig:hyper_lambda_1} and \ref{sfig:hyper_lambda_2}). Similar to the state of practice \cite{Geron2017}, some mild regularization terms help to improve prediction: the optimum $\lambda_1$ is about $0.01$ and the optimum $\lambda_2$ is about $0.0001$. (3) As to depth and width (Figures \ref{sfig:hyper_depth_shared}, \ref{sfig:hyper_depth_specific}, and \ref{sfig:hyper_width}), deeper and wider architectures do not appear to improve the performance. This part is consistent with our previous discussion that naively increasing model complexity does not necessarily improve model performance since larger model complexity leads to higher estimation errors. Also depth and width themselves do not reflect the speciality of MTL, as opposed to the more effective MTLDNN-specific architectural design, such as the ratio between shared and task-specific layers. Overall, it seems that the regularization methods specific to MTLDNNs, such as $\lambda_3$ and MTLDNN-specific architectural designs, are more important than these generic hyperparameters, at least in this one RP and SP dataset. 

\subsection{Interpreting MTLDNN for AV Adoption}
\noindent
MTLDNNs are not only predictive, but also interpretable. DNNs can be interpreted in at least two ways: visualization showing how choice probabilities change with input values, and importance ranking of input variables according to their elasticity values. Both methods are commonly used for interpreting DNN models \cite{Sung1998,Rao1998,Bentz2000,Baehrens2010,Simonyan2013}. 

\begin{figure}[htb]
\centering
\subfloat[AV Cost]{\includegraphics[width=0.3\linewidth]{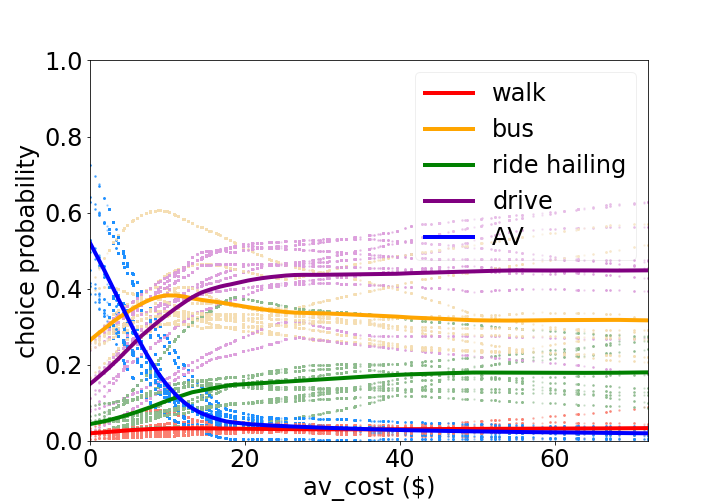}\label{sfig:av_cost}}
\subfloat[AV Wait Time]{\includegraphics[width=0.3\linewidth]{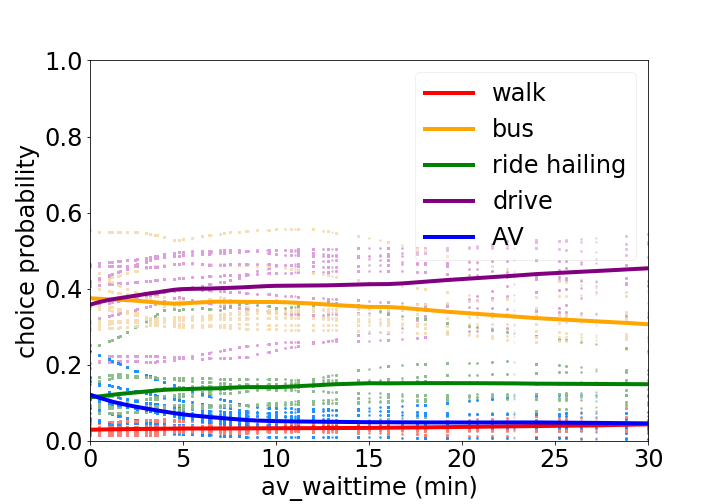}\label{sfig:av_wait_time}}
\subfloat[AV In-Vehicle Time]{\includegraphics[width=0.3\linewidth]{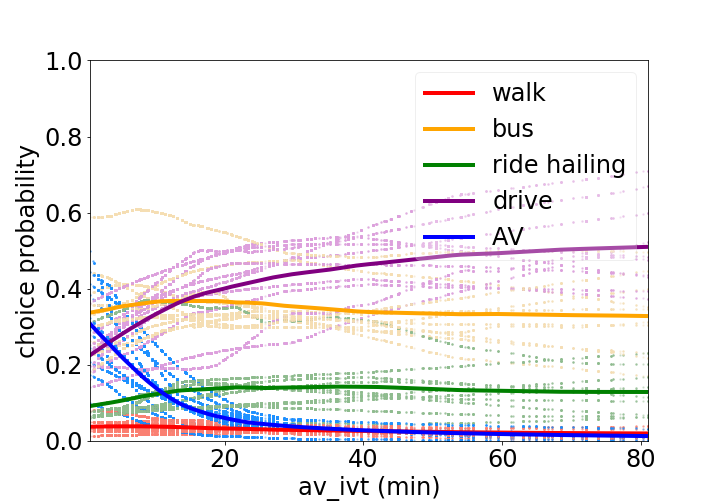}\label{sfig:av_in_vehicle_time}} \\
\subfloat[Age]{\includegraphics[width=0.3\linewidth]{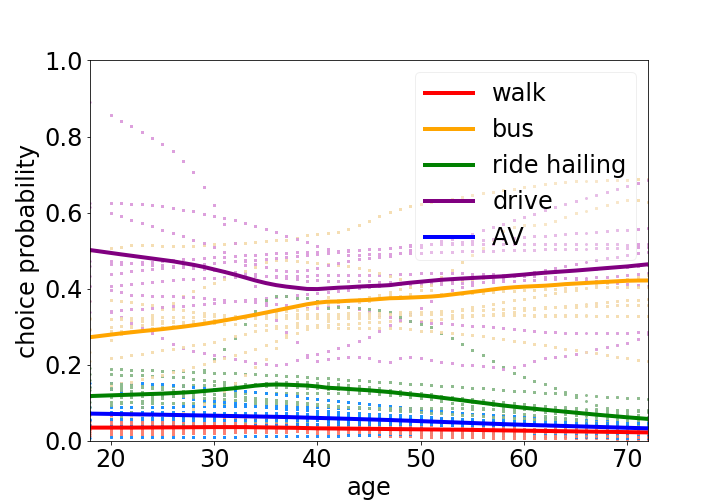}\label{sfig:av_age}}
\subfloat[Income]{\includegraphics[width=0.3\linewidth]{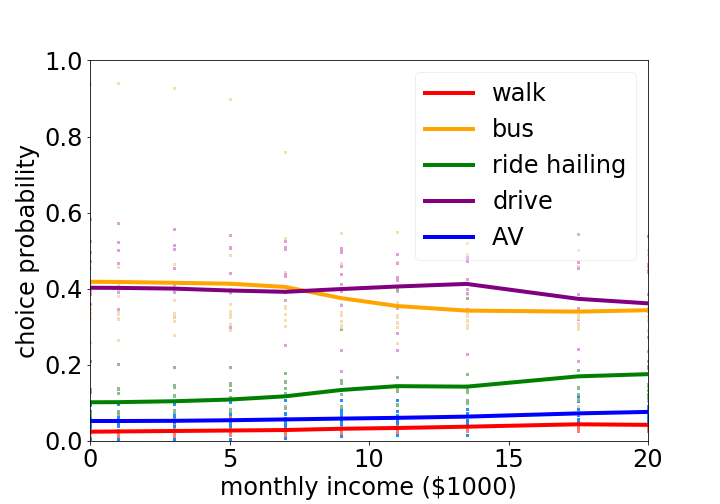}\label{sfig:av_income}}
\caption{Choice Probability Functions Varying with Inputs Values; light curves are the individual MTLDNN results; dark ones are the average of top 10 models.}
\label{fig:av_results}
\end{figure}

Figure \ref{fig:av_results} shows how the probabilities of choosing the five travel modes in the SP survey vary with AV costs, AV waiting time, AV in-vehicle travel time, age, and income. The first three variables are the most important alternative-specific variables, and the last two are important social-economic variables. AV is the specific focus since it is the targeting new technology that is designed into SP but does not exist in reality (RP). As shown by Figures \ref{sfig:av_cost}, \ref{sfig:av_wait_time}, and \ref{sfig:av_in_vehicle_time}, people are highly responsive to these AV-specific cost variables. For instance, the probability of choosing AV drops from about $50 \%$ to only $5 \%$, as AV cost increases from $\$ 0$ to $\$ 20$; Similarly, the probability drops from about $30 \%$ to only $5 \%$, as AV in-vehicle travel time increases from $0$ to $20$ minutes. Figures \ref{sfig:av_cost} and \ref{sfig:av_in_vehicle_time} also show that driving is the major substitute travel mode to AV. Relatively speaking, the probability of adopting AV is much less sensitive to socio-economic variables, as shown by Figures \ref{sfig:av_income} and \ref{sfig:av_age}: the probability curve of adopting AV is nearly flat everywhere, with respect to different values of age and income.

Table \ref{table:elasticity} presents the average elasticity of choice probability of AV with respect to input variables, sorted by the magnitude of the elasticity values. Different from Figure \ref{fig:av_results} that focuses on the visualization of the choice probability functions, Table \ref{table:elasticity} computes the elasticity values averaged over our sample, as a proxy to the elasticity of the population. It turns out that the results are not very different from those visualized in Figure \ref{fig:av_results}. One percent increase in AV cost and in-vehicle travel time leads to $0.981$ and $0.905$ percent decrease of the probability of choosing to use AV, while the impact of age and income is relatively smaller with values of $-0.561$ and $0.102$ respectively. These results suggest that methodologically it is feasible to extract reliable economic information from MTLDNN choice models and that substantially AV adoption still heavily depends on its cost structure, which is consistent with the long-standing results in travel behavioral analysis. 

\begin{table}[htb]
    \centering
    \begin{tabular}{@{}cccccc@{}}
        \toprule
        ﻿Variable & Elasticity \\
        \midrule
        AV Cost & -0.981 \\
        AV In-Vehicle Time & -0.905 \\
        Age & -0.561 \\
        AV Wait Time & -0.375 \\
        Income & 0.102 \\       
		\bottomrule
    \end{tabular}
    \caption {Average Elasticity of Choosing AV}
  	\label{table:elasticity}
\end{table}

\section{Conclusions and Discussions}
\noindent
This study introduces a MTLDNN framework to combine RP and SP for demand analysis. It is fueled by the practical importance of combining RP and SP for prediction and the theoretical interest of using deep learning models to answer classical questions. This study yields the following three major findings. 

First, it is theoretically feasible and effective to combine RP and SP datasets by using the MTLDNN framework, since it takes advantage of the automatic feature learning capacity in DNN and imposes soft and flexible constraints to capture the similarity and differences between tasks. MTLDNN is more generic than the classical method of using NL to combine RP and SP, due to the approximation power and the soft constraints, including diverse architectures and regularization methods, although MTLDNN could trigger larger estimation error due to its high model complexity. Second, MTLDNN empirically outperforms six benchmark models, and particularly outperforms the NL models with and without parameter constraints by about $5 \%$ prediction accuracy. This performance improvement is robust with respect to the choice of hyperparameters and architectural designs of MTLDNN models in both RP and SP datasets. This performance improvement can be mainly attributed to the DNN architectures and regularizations specific to the multitask learning problem, but not the generic approximation power of the standard feedforward deep architectures. Lastly, the results of MTLDNNs are interpretable by using gradient-based methods. We found that there exists a strong substitution pattern between driving and AV and that AV-specific variables (time and cost) play a more important role than socio-economic variables. 

It does not necessarily improve model performance by naively applying the standard DNN architectures, increasing the depth and width, and using default regularization methods without adjusting these factors to specific problems. This finding is partially an answer to the studies that find only limited improvement or no improvement of DNNs on classical choice models by directly applying standard DNN structures to travel mode analysis \cite{Nijkamp1996,Mozolin2000}. In fact for multitask learning, many new MTLDNN architectures are already created to capture the similarities and differences of the multiple tasks in a more subtle way than the architecture used in our study \cite{LongMingsheng2015,Hashimoto2016,Misra2016,Ruder2017_sluice}. Exploring these MTLDNN architectures for RP and SP is a promising future research direction. Researchers can also explore how to design new MTLDNN architectures specifically for RP and SP. Moreover, since the MTLDNN is just one particular DNN architecture, future studies can search for new MTLDNN architectures in an automatic way by using sequential modeling techniques \cite{Feurer2018,Vanschoren2018,Zoph2016,Zoph2017}.

\begin{figure}[htb]
\centering
\subfloat[MTLDNN Architecture]{\includegraphics[width=0.55\linewidth]{g_mtldnn.pdf}\label{sfig:mtldnn_2}}
\subfloat[Implicit Tree Structure in MTLDNN]{\includegraphics[width=0.25\linewidth]{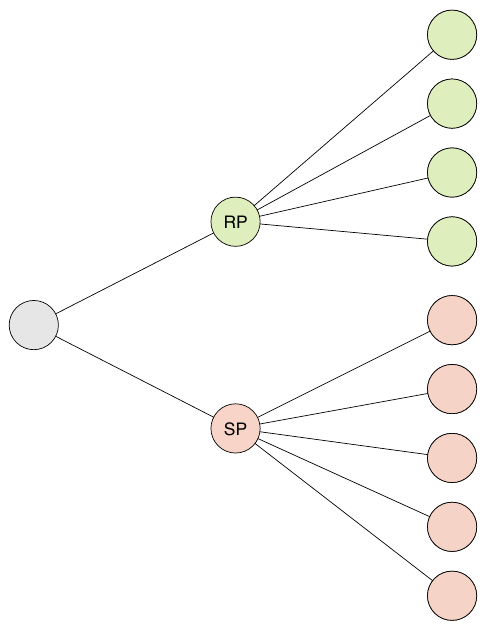}\label{sfig:tree}}
\caption{Shared Tree Structures in MTLDNN and NL}
\label{fig:tree}
\end{figure}

There is an intriguing relationship between MTLDNNs and NL, reflected in the similarity of their visualized architectures. Note that the meta-architecture of MTLDNN is exactly a tree-shaped structure (shown in Figure \ref{sfig:tree}), leading to our conjecture whether MTLDNNs can be used for \textit{all} the applications of NL model. However, in spite of their visual similarity, it is unclear how the mathematical formula of MTLDNN relates to that of NL, since the tree structure in MTLDNNs is the computational graph of input features, as opposed to that in NL reflecting the covariance matrix information of the random error terms. Nonetheless, this similarity is so intriguing that we encourage future studies to explore the deeper relationship between MTLDNNs and NL models.

While this study use the combination of RP and SP as one specific case, the MTLDNN is one generic method of combining different data sources. In the transportation domain, researchers have used simultaneous estimation (or structural equation models (SEM) in general) to jointly analyze different types of travel demands, such as auto ownership and mode choice \cite{Train1980,Zegras2010}, trip chains and travel modes \cite{YeXin2007}, travel time and vehicle miles traveled (VMT) \cite{Golob2003}; travel demand and attitudinal factors \cite{Lyon1984,Morikawa2002,Tardiff1977}; and activity patterns and travel demands \cite{Kitamura1992,Golob1997}. This MTLDNN framework can be easily applied to all these cases and extended to the scenarios in which the number of tasks is larger than two. Due to the power of MTLDNNs in combining RP and SP, it is likely for MTLDNNs to achieve a better performance in these cases as well.

Lastly, the application to RP and SP is not the most straightforward one for MTLDNNs, since the theoretical MTLDNN discussions often focus on only \textit{homogenous} cases. For instance, when MTLDNNs are used to combine the travel mode choices from two cities, the tasks are homogenous since the input and output dimensions of the two tasks match well. On the contrary, the tasks of RP and SP are \textit{heterogeneous} in the sense that the outputs of SP have one more alternative than RP and the inputs of SP have AV-specific variables that do not exist in RP. The remaining question even within our current framework is how to effectively engineer the input variables from the two tasks to improve the model performance. This is not a simple question since many different methods can apply \cite{GongJen2017}. Other unknown factors in this MTLDNN framework are caused by the generally insufficient research into the relationship between DNNs and statistical concerns \cite{Breiman2001}. For instance, RP and SP can be related as a panel structure since the unobserved random utility can be correlated; or RP and SP both have inherent preference heterogeneity across individuals \cite{Brownstone2000,Bhat2002}. A lot of statistical discussions involve the covariance structure of the random utility terms, which do not exist in DNNs at least in an obvious way. We believe it is essential to fully unleash the power of MTLDNNs if researchers can understand how to use DNNs to address statistical concerns such as heterogeneity and endogeneity. Simply put, there are infinite opportunities for researchers to use this MTLDNN for other types of applications and answer other theoretical questions. With the flexibility and the power of MTLDNN architectures, we believe that these further research directions will provide new insights into behavioral and policy analysis, and demand modeling methods.

\newpage
%\printbibliography
%\input{main.bbl}
\includepdf[pages={19,20,21,22,23}]{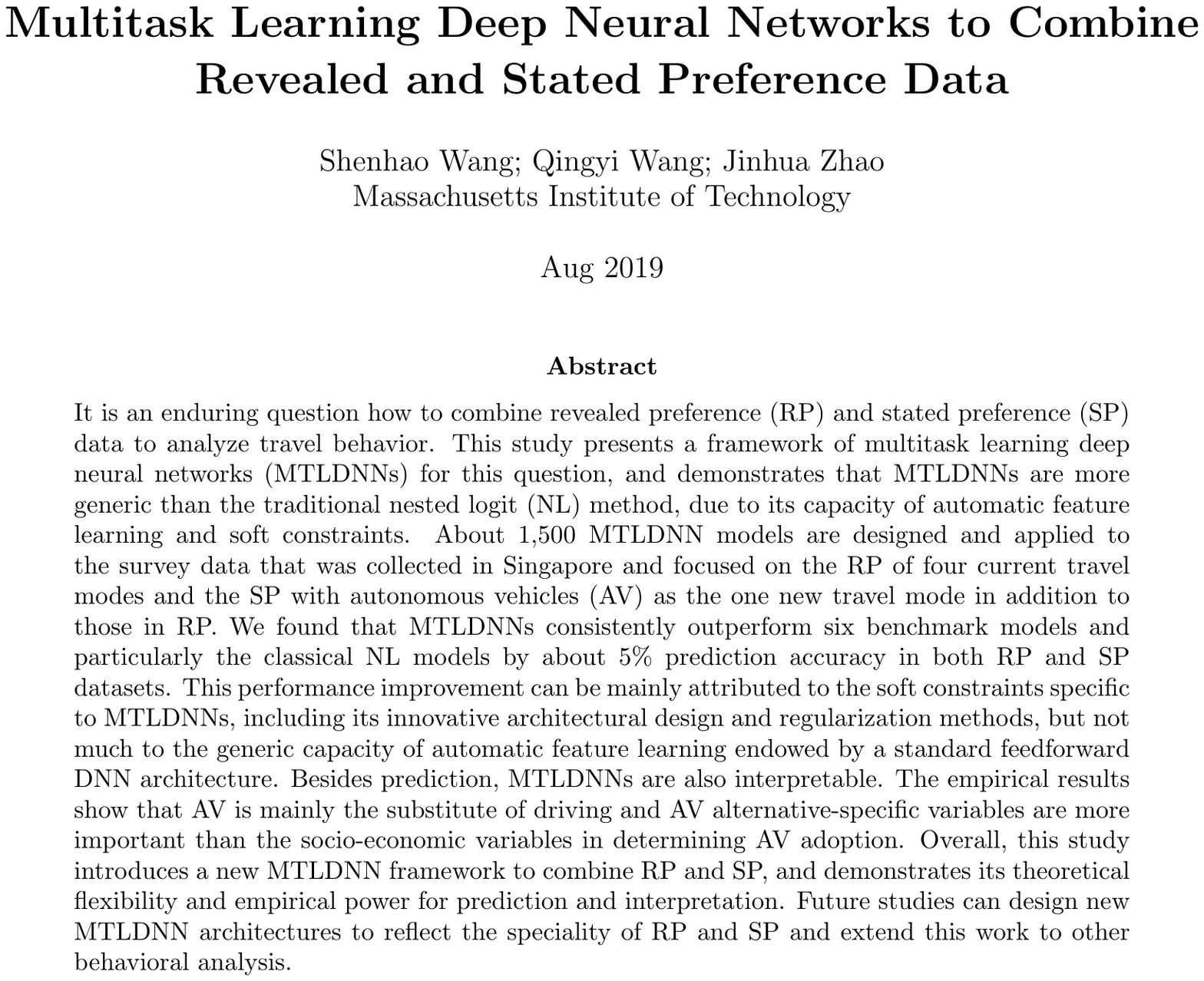}

\newpage
\section*{Appendix I: Hyperparameter Space}
\begin{table}[ht]
\centering
\resizebox{0.9\linewidth}{!}{%
\begin{tabular}{P{0.4\linewidth} P{0.6\linewidth}}
\hline
\textbf{Hyperparameter Dimensions} & \textbf{Values} \\
\hline
Shared M1 & $[1, 2, 3, 4, 5]$ \\
Domain-specific M2 & $[1, 2, 3, 4, 5]$ \\
$\lambda_1$ constant & $[1e{-20}, 1e{-4}, 1e{-2}, , 5e{-1}]$ \\
$\lambda_2$ constant & $[1e{-20}, 1e{-4}, 1e{-2}, , 5e{-1}]$ \\
$\lambda_3$ constant & $[1e{-20}, 1e{-4}, 1e{-2}, , 5e{-1}]$ \\
n hidden & $[25, 50, 100, 200]$ \\
n iteration & $20000$ \\
n mini batch & $200$ \\
\hline
\end{tabular}
} %resize used here
\caption{Hyperparameter Space of MTLDNN}
\label{table:arch_mtldnn}
\end{table}

\section*{Appendix II: Descriptive Summary Statistics}
\noindent
The age of the participants ranged between 20 and 85, and the income ranged from no income to over $20,000$ per month. A comparison of age and income distribution between the sample and the population is summarized in the following table. In terms of age, the sample overrepresents young people, and underrepresents the elderly. In terms of monthly income, individuals with no income and very high income more than $20,000$ are underrepresented in the sample, while the distribution of all other income groups was close to that of the population. All participants received monetary compensation for their responses.

\begin{table}[htb]
    \centering
    \begin{tabular}{@{}cccccc@{}}
        \toprule
        ﻿Age Group & Population (\%) & Sample (\%) & Income Group & Population (\%) & Sample (\%) \\
        \midrule
        $20-24$ & 8.42 & 16.31 & No income & 10.79 & 1.46 \\
        $25-29$ & 9.04 & 17.32 & Below \$2,000 & 7.49 & 7.19 \\
        $30-34$ & 9.22 & 15.45 & \$2,000 $-$ \$3,999 & 10.69 & 14.9 \\
        $35-39$ & 9.75 & 14.08 & \$4,000 $-$ \$5,999 & 11.29 & 17.35 \\
        $40-44$ & 10.12 & 10.09 & \$6,000 $-$ \$7,999 & 10.89 & 15.57 \\
        $45-49$ & 9.72 & 10.2 & \$8,000 $-$ \$9,999 & 9.49 & 14.77 \\
        $50-54$ & 10.19 & 7.42 & \$10,000 $-$ \$11,999 & 8.39 & 10.07 \\
        $55-59$ & 9.67 & 4.93 & \$12,000 $-$ \$14,999 & 9.09 & 8.22 \\
        $60-64$ & 8.13 & 2.49 & \$15,000 $-$ \$19,999 & 9.49 & 4.78 \\
        $65-69$ & 6.39 & 0.67 & Over \$20,000 & 12.39 & 5.69 \\
        $70-74$ & 3.35 & 0.91 &  &  &  \\
        $75-79$ & 2.84 & 0 &  &  &  \\
        $80-84$ & 1.73 & 0.13 &  &  &  \\
        $85+$ & 1.43 & 0 &  &  &  \\
        \bottomrule
    \end{tabular}
    \caption{Survey Descriptive Summary Statistics}
\end{table}

\section*{Appendix III: Top 10 MTLDNN Architectures}
\begin{table}[htb]
    \centering
    \begin{tabular}{@{}cccccc@{}}
        \toprule
        ﻿Shared M1 & Domain-specific M2 & n hidden & $\lambda_1$ & $\lambda_2$ & $\lambda_3$ \\
        \midrule
		1&1&25&1.00E-02&1.00E-02&1.00E-04\\
		3&2&25&1.00E-02&1.00E-04&1.00E-20\\
		1&1&25&1.00E-20&1.00E-02&1.00E-02\\
		1&1&25&1.00E-02&5.00E-01&1.00E-04\\
		1&1&100&1.00E-02&1.00E-20&1.00E-04\\
		1&4&25&1.00E-02&5.00E-01&1.00E-02\\
		1&1&200&1.00E-02&5.00E-01&1.00E-02\\
		1&1&50&1.00E-02&1.00E-02&1.00E-20\\
		3&1&100&1.00E-02&1.00E-04&1.00E-04\\
		2&3&50&1.00E-02&5.00E-01&1.00E-20\\      
		\bottomrule
    \end{tabular}
    \caption{Top 10 MTLDNN Architectures}
\end{table}

\end{document}